\documentclass[preprint,12pt]{elsarticle}

\usepackage{amssymb}
\usepackage{float}
\usepackage{amsmath}

\journal{Journal of Fluids and Structures}

\begin{document}

\begin{frontmatter}



\title{Vibrissa inspired geometries enhance sensitivity of wake-induced vibrations}


\author{Eva Erickson} 
\affiliation{organization={Center for Fluid Mechanics, School of Engineering},
            addressline={}, 
            city={Providence},
            postcode={02912}, 
            state={RI},
            country={USA}}
\author{Eric E. Handy-Cardenas} 
\affiliation{organization={Center for Fluid Mechanics, School of Engineering},
            addressline={}, 
            city={Providence},
            postcode={02912}, 
            state={RI},
            country={USA}}
\author{Joel W. Newbolt} 
\affiliation{organization={Center for Fluid Mechanics, School of Engineering},
            addressline={}, 
            city={Providence},
            postcode={02912}, 
            state={RI},
            country={USA}}
\author{Christin Murphy} 
\affiliation{organization={Naval Underwater Warfare Center},
            addressline={}, 
            city={Newport},
            postcode={02841}, 
            state={RI},
            country={USA}}
\author{Kenneth Breuer} 
\affiliation{organization={Center for Fluid Mechanics, School of Engineering},
            addressline={}, 
            city={Providence},
            postcode={02912}, 
            state={RI},
            country={USA}}
\begin{abstract}
We report on experiments designed to characterize the vortex-induced vibration (VIV) and wake-induced vibration (WIV) experienced by bluff bodies immersed in both steady and unsteady flows. Using a real-time Cyber-Physical System (CPS) we systematically prescribe the virtual mass, spring constant, and damping of elastically mounted models. This allows us to characterize the forces and displacements of the free vibration of a circular cylinder, elliptical cylinder, and a seal whisker inspired vibrissa model with undulating elliptical geometry. In a free flow, the circular cylinder exhibits high VIV, while the reduced aspect ratio objects have minimal vibration across all structural frequencies. When a flow disturbance of a pitching and heaving hydrofoil is introduced, the reduced aspect ratio objects are excited by WIV with highest amplitude oscillations occurring when structural frequency of the test object matches wake frequency of the upstream foil. To further understand the benefits of an undulated geometry over a classic elliptical cylinder, we assess the nonlinear fluid damping experienced by each test object by comparing experimental data to quadratic drag and Van der Pol damping models. Our results show that the amplitude dependent Van der Pol damping model better describes the physical system for both test objects by capturing the suppression of large amplitude WIV, but recovering small amplitude VIV. However, the strength of the fitted Van der Pol damping coefficient is greater for the elliptical cylinder than the vibrissa. We find the vibrissa experiences lower damping than the elliptical cylinder across all tested structural frequencies, indicating how the vibrissa geometry may serve as a higher sensitivity sensor.  
\end{abstract}

\begin{highlights}
\item Cyber–physical system enables systematic study of vortex- and wake-induced vibration
\item Vibrissa-shaped objects suppress VIV and amplify WIV
\item Damping coefficient assessed for Van der Pol and quadratic damping models 
\item Van der Pol damping model captures nonlinear fluid damping experienced
\item Vibrissa model experiences less fluid damping, enhances sensing sensitivity
\end{highlights}


\end{frontmatter}



\section{Introduction}
Vortex-induced vibrations (VIV) and wake-induced vibrations (WIV) are key phenomena in understanding the oscillation of an elastically mounted bluff body in a flow. VIV are induced by unsteady vortex shedding from an objects' own body \citep{williamson2004vortex}. WIV, in contrast, occur when the wake developed from an upstream object drives the vibration of a downstream flexible body. This dynamic response of the downstream body due to the characteristics of the upstream wake is also referred to in the literature as wake-induced gallop or interference galloping \citep{bokaian1984wake,hover2001galloping}. The VIV and WIV responses of bluff bodies in a flow have been extensively investigated in the last 50 years. The case of a rigid circular cylinder being elastically mounted and constrained to oscillate transversely to a freestream has been well-studied, \citep{sarpkaya1979vortex,bearman1984vortex,parkinson1989phenomena,bahmani2010effects}. Particularly influential studies regarding the VIV dynamics of this system with low structural damping and a low damping ratio come from \citet{khalak1996dynamics, khalak1997fluid, khalak1997investigation, khalak1999motions}, who assess how the amplitude of VIV of an elastically mounted circular cylinder with a low damping ratio changes over a range of reduced velocity $U^* =U/f_nD$, where $U$ is freestream flow velocity, $D$ is diameter of the cylinder, and $f_n$ is the natural (structural) frequency. If the vortex shedding frequency is near $f_n$, a nonlinear resonance phenomenon known as ``lock-in'' can be encountered with large-amplitude body oscillations. A hydroelastic experimental set-up allowed them to adjust the stiffness and damping ratio of the system by modifying physical springs to incite this phenomenon. Sweeps of $U^*$ were conducted by changing the flow velocity, which also changed the Reynolds number of the flow. Results show lock-in excitation in cylinder vibration occurs for a mid-range $U^*$ with a secondary higher amplitude region, termed the ``upper branch'' for a small range of $U^*$ around 3 to 6. The relative range of excitation of the lock-in branches have a damping ratio dependence, with access to the higher excitation region coming from reducing the damping ratio of the system. 


To understand the fluid damping experienced during VIV, different modeling techniques have been proposed. Mathematical modeling of damping experienced by bodies moving through fluid has long been described by a quadratic drag model, where resistive force is proportional to the square of relative velocity \citep{smith1998newton}. \citet{lin2009nonlinear} examined the role of quadratic drag in VIV and found that nonlinear fluid damping in the form of a velocity squared term can be applied as an independent fluid effect for the vortex lift and added-fluid force. In addition to quadratic drag, a Van der Pol wake oscillator coupled with a spring–mass–damper system has been proposed by \citet{facchinetti2004coupling} to model the complex damping involved in vortex- and wake-induced vibration systems. A classical Van der Pol equation is used to model the near wake dynamics describing the fluctuating nature of vortex shedding and is coupled to the structural dynamics of a circular cylinder. This allowed them to reproduce key VIV phenomena such as lock-in and amplitude responses, marking a benchmark in understanding VIV of circular cylinders. 

Modifying the aspect ratio (AR = minor axis / major axis) of elliptical cylinders induces different VIV than a circular cylinder with $AR = 1$. \citet{zhao2019dynamic} characterized the vibration of elastically-mounted elliptical cylinders with aspect ratios $0.67 \leq AR \leq 1.50$ in a free flow and found that the amplitude of excitation has a strong aspect ratio dependence 
and showed that reduced aspect ratio objects experienced only two lock-in regions, while those with $AR \geq 1$ experienced three and displayed higher amplitude vibration. 
One factor contributing to the VIV reduction in elliptical cylinders is the modification of the separation point. \citet{CAGNEY2019316} 
calculated the separation angle on a circular cylinder throughout the shedding cycle for various wake modes, and 
found that all wake modes associated with VIV had a periodic movement of the separation point. However, for elliptical cylinder models, there was negligible movement in the separation point, which suppressed VIV. Overall, bluff bodies with reduced aspect ratio (minor axis $<$ major axis) demonstrate VIV suppression in a flow across reduced frequencies.

An example of this phenomenon occurring in nature comes from the sensory system of harbor seals and other members of the \textit{phocidae} family. The amazing navigation and prey-tracking abilities of seals are enabled by the unique structure of their facial whiskers called vibrissa. Vibrissa are known to have an elliptical cylinder shape with a three-dimensional undulating geometry \citep{zheng2023wavy,wang2016wake, geng2024single}. An instrumental study by \citet{hanke2010harbor} investigated the hydrodynamic function of this undulated structure 
and found that the dynamic forces on harbor seal whiskers are, by at least an order of magnitude, lower than those on non-undulated sea lion whiskers. They concluded that by reducing vibrations, seals keep their vibrissa as still as possible while detecting relevant hydrodynamic signals, thus enhancing the signal-to-noise ratio. However, their work did not explicitly test this hypothesis. \citet{lyons2020flow} dove deeper into understanding the fluid flow effects from specific geometric features of vibrissa in a computational investigation of seal whisker morphological parameters including wavelength, aspect ratio, undulation amplitudes, symmetry, and undulation off-set. At Re=500, large-eddy simulations were computed for the flow over 16 different (stationary) seal whisker topographies with geometric modifications derived from the model proposed by \citet{hanke2010harbor}. In agreement with aforementioned elliptical cylinder literature, aspect ratio ranked as highest impact parameter. Modifying the magnitudes of both major (streamwise) and minor (transverse) axis undulations was also proven critical to force and frequency responses, respectively. 
Anechoic wind tunnel experiments at much higher Reynolds number (Re=60k) by \citet{zhu2024flow} further emphasized the important three-dimensional effects of undulation in a vibrissa on the flow field showing that three-dimensional flow separations introduced by the alternative saddle and nodal planes of the seal vibrissa model inhibit shear layer interactions and reduce the spanwise coherence. The result is a wake with no dominant coherent structures, thus suppressing pressure fluctuations on the solid surfaces. A review by \citet{lekkala2022recent} provided a very comprehensive comparison of the wake formations for even more unique cylindrical bluff bodies cross-sections including helically twisted elliptical, symmetric wavy, asymmetric wavy, and seal vibrissa geometries. They summarized that Reynolds number and geometry are two prominent parameters that govern the formation and behavior of the flow over a bluff body, with Reynolds number becoming increasingly dominant over geometry. Overall, vortex-shedding suppression is greatly affected by undulation and aspect ratio, but as flow velocity or angle of attack increases, the flow features tend towards those of a classic circular cylinder.

Much work has been done on understanding the WIV response of classical geometries to vortical wakes, however studies of the vibrissa geometry have been largely focused on the clean flow case. A study by \citet{beem2015wake} assessed the WIV response of vibrissa by positioning an elastically mounted test model different distances downstream in the hydrodynamic wake of stationary upstream vortex shedding circular cylinders 
The flexibly mounted vibrissa vibrated freely with large amplitude, with a frequency matching the shedding frequency of the upstream cylinder, and following a path that slaloms among the vortices of the oncoming wake. 

While this extensive prior research has addressed many fundamental issues regarding vibration phenomena for systems of differing mass and damping, there are still many questions presented at the intersection of physical and simulated experimentation, especially regarding models deviating from the classical geometry of a circular cylinder. In the present work, we combine simulation and physical experimentation using a Cyber-Physical System (CPS) as a tool to implement simulated parameters to physical elastically mounted structures with real-time force feedback. With the CPS, we can systematically canvas dynamical systems on a physical stage across reduced velocities and structural frequencies without changing the flow speed, the Reynolds number, or physically changing the spring and damping components. Our experimental set-up provides an efficient and effective way to independently vary the structural frequency. 

In this work, we characterize both the vortex- and wake-induced vibration responses of three different elastically mounted bluff body models: circular cylinder, elliptical cylinder, and a seal vibrissa-inspired undulated elliptical cylinder model. Specifically, we test the hypothesis that the vibrissa-shaped bodies found in nature are more sensitive to wake disturbances than smooth low-aspect ratio geometries. The CPS also allows us to explicitly test the VIV response as a function of Reynolds number - a study not possible when one controls $U^*$ by changing the flow speed, with a fixed value of $f_p$. Lastly, we can assess the ability of different damping models -  proven to describe the nonlinear damping in circular cylinder vibration systems -  to accurately capture the damping experienced by  elliptical and seal-whisker inspired geometries. 

In the following sections, we first validate the CPS by comparing free flow VIV results (displacement and force) to those found in the literature. We then introduce upstream vortical disturbances, generated by a pitching and heaving hydrofoil, and characterize the resulting WIV of the downstream test models. Finally, we compare the nonlinear damping experienced by reduced aspect ratio objects to gain greater insight to the contributions of an undulated geometry to vibrational performance. 

\section{Methods}
\begin{figure}[H]
    \includegraphics[width=10cm]{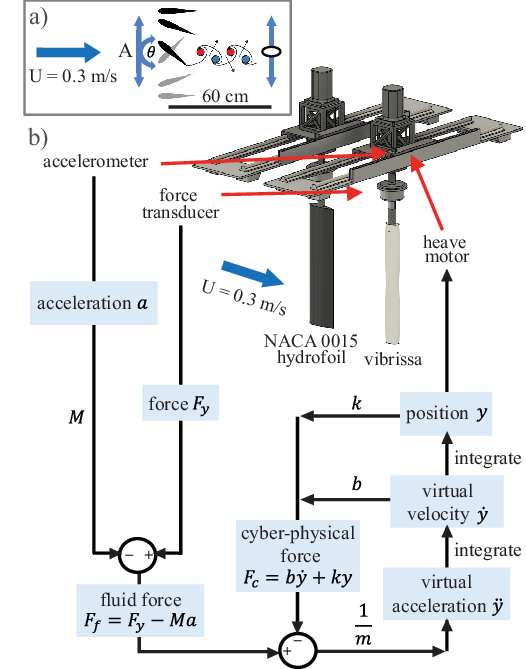}
    \centering
    \caption{\textbf{Experimental set-up.} (a) Top view of the experimental set-up. (b) Schematic demonstrating the connection between the physical components of the downstream-mounted body and the cyber-physical system governing its motion. Values $k$, $b$, and $m$ are the commanded spring constant, damping constant, and mass, respectively, while $M$ is true physical mass. 
    }
    \label{fig:set up}
\end{figure}
All experiments are conducted in the Brown University free-surface water tunnel with a test section of width × depth × length = 0.8 m × 0.6 m × 4.0 m. Figure~\ref{fig:set up} shows a schematic of the experimental set-up. A NACA0015 hydrofoil with a 12 cm chord and 36 cm span is mounted vertically on a computer-controlled gantry and oscillated in a pitch-heave motion so as to generate a propulsive vortex wake \citep{chao2021drag}.  The test article is mounted on a second gantry, positioned 60 cm ($5c$) downstream measured center to center.  Three models were tested: a circular cylinder (length × diameter = 0.450 m × 0.054 m), elliptical cylinder (length × major axis × minor axis = 0.405 m × 0.054 m × 0.024 m), and seal vibrissa model (length = 0.405 m). The seal vibrissa model has two coordinated sets of spanwise undulations along the $z$-axis designed based on geometry adopted by \citet{lyons2020flow}. This geometry is formed from two different sized ellipses a distance $S$ apart, with major radii $a$ and $k$, and minor radii $b$ and $l$. Where the ellipse major-axis is at its maximum value, the ellipse minor-axis is at its minimum value and vise versa. Our models include two undulations with values of $S$ = 0.100 m, $a$ = 0.060 m, $b$ = 0.024 m, $k$ = 0.047 m, $l$ = 0.029 m. All three test articles had the same hydraulic diameter, $D = 0.054$m.  Experiments were conducted at a freestream velocity of $U=0.3$ m/s, yielding a Reynolds number Re$=\rho U D/\mu=16200$, where  $\rho$ and  $\mu$ are water density and dynamic viscosity, respectively. A limited study of Reynolds number sensitivity was also performance by repeating select experiments at a different freestream velocity. 

The test model is mounted on a six-axis force/torque transducer (ATI 9105-TIF-Delta-IP65), which in turn is mounted on a gantry that is free to heave in the $y$-direction, controlled by a linear motor.  
This transverse motion is controlled by a real-time Cyber-Physical System (CPS) which emulates a flexible mount with a prescribed stiffness and damping \citep{hover1998forces,mackowski2011developing,onoue2015large,zhu2021nonlinear}.  The forces on the vibrissa, $\mathbf{F}(t)$, along with the heave position, $y(t)$, and a direct measurement of the heave acceleration, $\ddot y(t)$  (PCB Piezotrontics differential MEMS DC accelerometer) are used with a real-time feedback control computer to simulate a flexible mount of the model with arbitrary values for the stiffness, $k$, damping, $b$ and mass, $m$. This provides a simple and fast way to change the characteristics of the structural support and thus to explore a wide range of dynamics of the WIV system. A schematic of the CPS is shown in Fig~\ref{fig:set up}, and full details are described in \citet{zhu2020nonlinear}.  

For this study the mass ratio, $m^*$ - the ratio of the object density to the water density - was fixed at 10.8. The stiffness, $k$ changed so as to vary the structural frequency $f_s = \sqrt{k/m}/(2\pi)$. The structural damping, $c$ was adjusted to maintain a constant structural damping ratio $\zeta_s=b/(2\sqrt{mk})=0.003$. Note that while this represents a very lightly damped oscillator, the fluid forces provide significant damping and so the structural damping is effectively negligible.

Adjusting the structural frequency, while keeping the freestream velocity fixed allows us to sweep through a range of reduced velocities, $U^*=U/D f_s$ from 2 to 11, while keeping the Reynolds number, Re$ = \rho U D/ \mu$ fixed at 16200.  This is distinct from most previous experiments on VIV and WIV in which the flow speed is used to control $U^*$. 

Reduced velocity sweeps were conducted for both clean flow (VIV) and disturbed flow (WIV) configurations.  It is important to emphasize that the motion of the test model is never commanded, but is free to oscillate, through the CPS-prescribed mass, spring and damping constants. 

To generate the vortical wake, the upstream hydrofoil was moved with a sinusoidal pitching and heaving motion to generate an unsteady ``thrust wake'' consisting of alternating vortices \citep{chao2021drag}. Six wake frequencies were used: $f_w$=1.39, 1.01, 0.93, 0.79, 0.69 and 0.56 Hz, with a fixed pitch amplitude of 20 degrees.  The heave amplitude, $A$, was varied so that the Strouhal number $St=A f_w /U$ was held constant at a value of 0.2, corresponding to the range observed for swimming fish \citep{eloy2012optimal}.
At each value of the parameters $f_w$ and $f_s$, the test model was released from a $y$-displacement of zero and allowed to equilibrate for 90 cycles to ensure that all transients had died out and a steady state condition had been achieved. Following this, data was collected for 140 cycles. Here, one cycle was defined by the structural frequency, $f_s$, for free flow experiments and by the wake frequency, $f_w$, for disturbed flow experiments.

\section{Results and discussion}
\subsection{Vortex-induced vibration}
\begin{figure}[H]
    \includegraphics{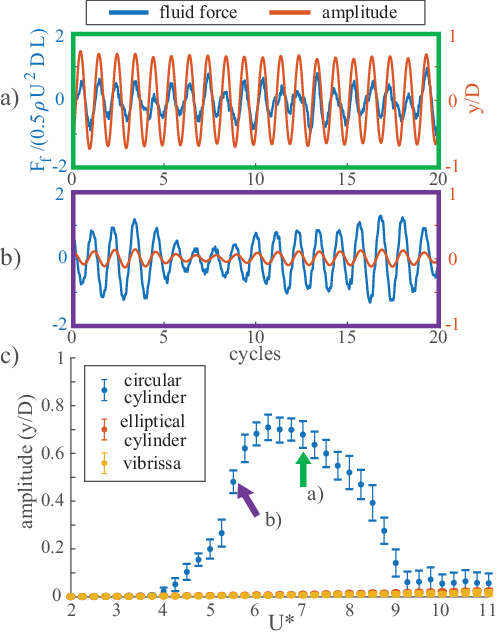}
    \centering
    \caption{\textbf{Vortex-induced vibration response of test models in free flow.} Y-axis displacement and fluid force measurements plotted together over 20 cycles for circular cylinder trials conducted at (a) $U^*=7$ and (b) $U^*=4.5$. (c) Complete $U^*$ sweeps from $U^*=2$ to $U^*=11$ for circular cylinder, elliptical cylinder and vibrissa models.  
    }
    \label{fig:free flow}
\end{figure}

The vortex-induced vibrations for a circular cylinder in a clean flow agree with results found in similar experiments using physical springs \citep{khalak1996dynamics,feng1968measurement,brika1993vortex,liang2018viv}. Time series data (Figure~\ref{fig:free flow}(a-(b)) was collected for a range of reduced velocities and summarized by plotting the mean and standard deviation of the displacement maxima over all recorded time cycles at each $U^*$ (Figure~\ref{fig:free flow}(c)). We observe large scale oscillations for $U^*$ values ranging from 4 to 9 with peak oscillation amplitudes between $U^*=$ 6 and 7. We do not see the multi-branch regions documented by \citet{khalak1997fluid}, despite having similar mass-damping ratios ($m^*\zeta = $ mass ratio × damping ratio). Our non-dimensional amplitudes at peak excitation are roughly 0.7 which falls between their highest values in the lower branch of 0.6 and their lowest values in the upper branch of 0.8. We believe that our experimental procedure, which allows the VIV to evolve naturally from $y = 0$, only accesses the lower branch regime. We did not attempt to deliberately access the high amplitude branch. At some values of $U^*$, we observe a bi-stable response, where the amplitude of oscillation switched between larger and smaller amplitude modes. This presents itself as a beating phenomena (Figure~\ref{fig:free flow} (b)). 

In sharp contrast to the VIV response of the circular cylinder in a clean flow, the elliptical cylinder and vibrissa geometries exhibit little to no vibration across all reduced velocities. (Figure~\ref{fig:free flow} (c)). This agrees with the significant reduction in the periodic forces inducing minimal VIV on vibrissa compared to circular cylinders found in PIV and CFD experiments on vortex shedding of fixed seal whisker models by \citet{hanke2010harbor}, as well as results from free-vibration studies done on elastically mounted smooth or undulated elliptical cylinders by \citet{zhao2019dynamic, CAGNEY2019316, beem2015wake}.

\subsection{Reynolds number dependence}
\begin{figure}[H]
     \includegraphics{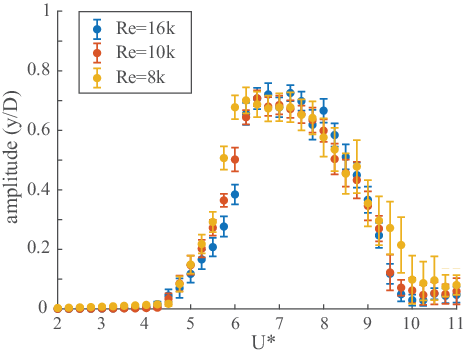}
  \centering
    \caption{\textbf{Vortex-induced vibration response of the circular cylinder is independent from Reynolds number}. $U^*$ sweep results for the circular cylinder at Re  $\sim$ 16k, 10k and 8k.}
    \label{fig:Re_dependence}
 \end{figure}

Many notable contributions to the study of vortex- and wake-induced vibrations have assessed excitation amplitude over a range of reduced velocities. In other works, access to a range of $U^*$ has been achieved by changing the flow speed, $U$. In doing so, the Reynolds number changes as well. To assess Reynolds number dependence, we repeated the free flow VIV experiments with the circular cylinder at three flow speeds: 15 cm/s, 20 cm/s, and 30 cm/s, corresponding to Re $\sim$ 8k, 10k, and 16k, respectively. In Figure~\ref{fig:Re_dependence} we see that the VIV amplitude response appears minimally affected by Reynolds number. The $U^*$ range of excitation is slightly broader at lower Reynolds number with the onset of lock-in occurring roughly a quarter $U^*$ earlier for Re=8k compared to Re=16k, however the amplitudes observed for all tested Reynolds numbers maintain the same trend. We conclude that while our experimental method is distinct from those prior in our ability to maintain constant Reynolds while varying $U^*$, VIV is independent of Reynolds number in this Reynolds number regime; thus controlling $U^*$ by varying the flow speed should not affect the results, as long as the range of velocities remains modest.
 
\subsection{Wake-induced vibration}
\begin{figure}[H]
    \includegraphics{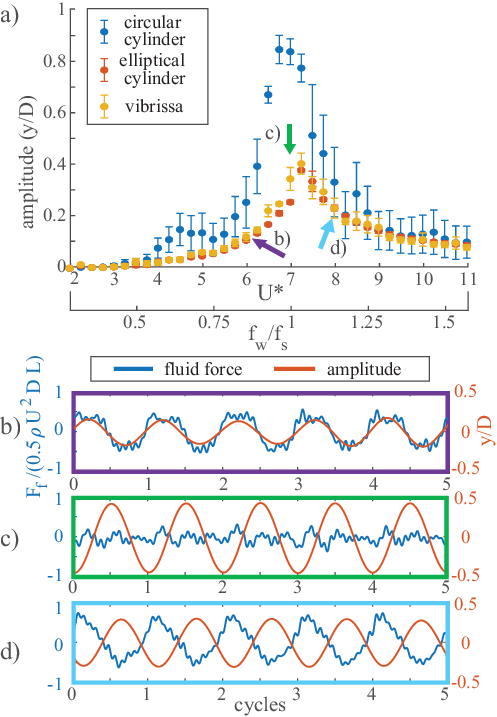}
    \centering
    \caption{\textbf{Wake-induced vibration response of test models in a vortical thrust wake.} (a) Complete $U^*$ sweeps from $U^*=2$ to $U^*=11$ for circular cylinder, elliptical cylinder and vibrissa models with an upstream wake frequency $f_w=0.79$ Hz. Y-axis displacement and fluid force measurements plotted together over 5 cycles for vibrissa trials with an upstream wake frequency $f_w=0.79$ Hz such that $f_w/f_s=1$ at $U^*=7$, conducted at (b) $U^*=6$ ($f_w/f_s<1$), (c) $U^*=7$ ($f_w/f_s=1$), and (d) $U^*=8$ ($f_w/f_s>1$).}
    \label{fig:disturbance}
\end{figure}

With the upstream hydrofoil in motion, the test articles are exposed to a vortex wake with a specific frequency, $f_w$. Figure~\ref{fig:disturbance} (a) shows the amplitude response over a range of $U^*$ of the circular cylinder, elliptical cylinder and vibrissa models for the case of $f_w = 0.79$ Hz. At this frequency, $f_w/f_s=1$ at $U^*=7$. As before, the symbols depict the average maximum vibrational amplitude, and the error bars represent the standard deviation of the maxima, computed from 140 cycles (after reaching steady state). The response of the circular cylinder (blue symbols) resembles a more variable version of the free flow case (Fig~\ref{fig:free flow}, blue symbols): for mid-range reduced velocities, the circular cylinder exhibits increased amplitudes of oscillation. The standard deviation of the mean maximum amplitudes of oscillation is much larger throughout the $U^*$ sweep. For the reduced aspect ratio objects - the elliptical cylinder and the vibrissa (red and yellow symbols respectively)  - the vibrational response greatly changes due to the presence of the upstream flow disturbance.

Figure~\ref{fig:disturbance} (b), (c), and (d) show 5 cycles of the vibrissa response at  $U^* = 6, 7,$ and 8 respectively. In the clean flow, vibrissa and elliptical cylinder VIV is almost negligible. In sharp contrast, when the model encounters an oncoming vortical wake, it vibrates with a frequency matching the flapping frequency of the upstream foil, $f_w$. This is demonstrated for a disturbance with $f_w=0.79$ Hz in Figure~\ref{fig:disturbance} (b-d) as amplitude of oscillation is plotted per cycle defined by wake frequency, for 3 different structural frequencies. For structural frequencies tested, the model oscillates at the wake frequency of the upstream disturbance. When the wake frequency is equal to the structural frequency of the test model, $f_w/f_s = 1$, the reduced aspect ratio objects exhibit their highest amplitude of oscillation (Figure~\ref{fig:disturbance} (a), yellow and orange symbols). At $U^*=6$, where $f_w/f_s<1$ (Figure~\ref{fig:disturbance} (b)), fluid force and amplitude are in phase with each other.  Conversely, when $U^*=8$, where $f_w/f_s>1$ (Figure~\ref{fig:disturbance} (d)), the fluid force and amplitude are out of phase with each other. When the structural frequency of the vibrissa and the wake frequency of the upstream foil are equal, $f_w/f_s=1, U^*=7$ (Figure~\ref{fig:disturbance} (c)), the amplitude of oscillation is at its maximum, and the fluid force appears to be at a minimum with only small fluctuations near zero. This behavior - the amplitude and force phase difference switching from zero to $180^\circ$ is the classic signature of forced response of a damped second order system as it moves through its natural resonant frequency - a response that is not surprising since the vibrissa is a spring-mass-damper system with the upstream foil simply serving as a harmonic forcing function.

Our results agree with the findings of \citet{beem2015wake}: the WIV of the reduced aspect ratio objects is substantially higher than the free flow case and the frequency of oscillation is tied to the frequency of incoming vortices shed from the upstream object. A difference however, is that the disturbance used by \citet{beem2015wake} was a stationary circular cylinder, generating a drag wake, while the flow disturbances in our experiments were produced by a pitching and heaving hydrofoil, generating thrust wakes analogous to those shed by swimming of fish or other prey.

\subsection{Comparison of fluid damping metrics}
A puzzling feature raised by these results is the observation that the elliptical cylinder and the vibrissa have very similar responses, both in the clean and the forced flow. 
This begs the question: if they have the same VIV and WIV response, what, if any, might be the dynamical function and/or benefit of the vibrissa's undulating geometry exhibited in nature? 

\begin{figure}[H]
    \includegraphics[width=\textwidth]{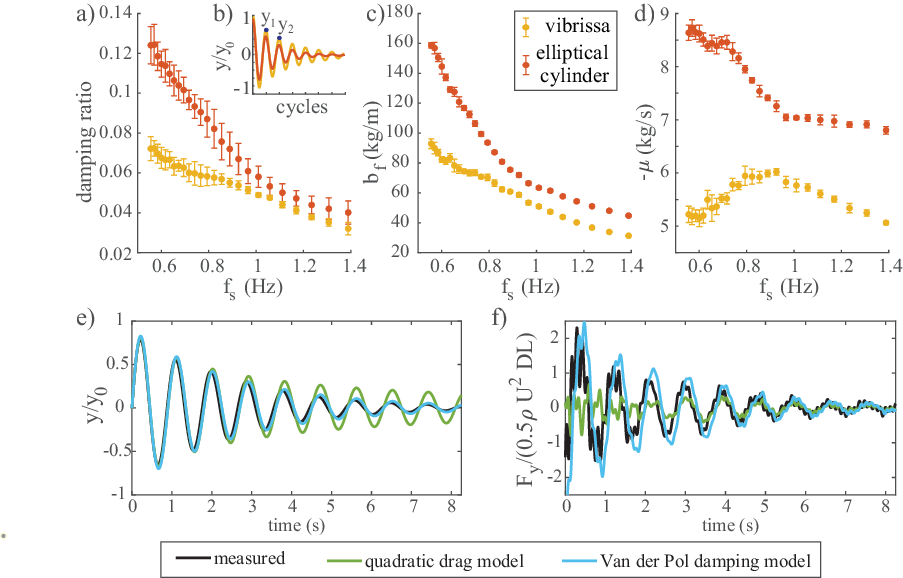}
    \centering
    \caption{\textbf{Comparison of damping metrics in ringdown experiments.} (a) Mean and standard deviation of damping ratios calculated using the method of logarithmic decrement by Equation \ref{eq:logdec} from ringdown trials at each tested structural frequency. (b) An example of a single ringdown experiment for both models at $f_s=0.79$ Hz. (c) Mean and standard deviation of quadratic drag coefficient $b_f$ calculated by optimization of damping equation \ref{eq:quad}. (d) Mean and standard deviation of Van der Pol damping coefficient $\mu$ calculated by optimization of damping equation \ref{eq:vdp}. Predicted and measured (e) position and (f) force for both damping models for $f_s=1.1$ Hz.}
    \label{fig:dampingratio}
\end{figure}

To address this, we calculate the damping ratio, $\zeta$, of the vibrational system.  Recall that the structural damping of the model was fixed with a value of $\zeta_s = 0.003$.  However, the observed damping is much higher, dominated by fluid damping, associated with the vortex shedding from the structure \citep{zhu2021nonlinear}. To quantify and compare the fluid damping experienced by the elliptical cylinder and vibrissa geometries we conducted ``ringdown'' experiments in which the test object was positioned in a freestream of 0.3 m/s, displaced 3 cm from its equilibrium position and released, allowing it to return to $y  = 0$, in a  motion characteristic of a damped harmonic oscillator. These experiments were repeated over a range of structural frequencies from 0.55 - 1.4 Hz, and for each tested structural frequency, 5 ringdown tests were conducted. For each ringdown, analysis was conducted on the measured $y$-force and $y$-position from the test object's first crossing of the zero position after being released from the 3 cm displacement, to the first peak in position at which displacement is less than 0.08 cm. Damping ratios are calculated using logarithmic decrement such that

\begin{equation}
    \zeta=\frac{\ln{(\frac{y_1}{y_2}})}{\sqrt{4\pi^2+\ln{(\frac{y_1}{y_2}})}}
    \label{eq:logdec}
\end{equation}

where $y_1$ is the amplitude of one peak at time $t_1$, and $y_2$ is the amplitude at the next peak, one cycle or $T$ seconds later, $y_1=y(t_2)$ and $t_2 = t_1+T$. Figure ~\ref{fig:dampingratio} (b) demonstrates an example of a ringdown at a structural frequency $f_s=0.79$ Hz with $y_1$ and $y_2$ labeled for the first and second peaks of the ringdown. This calculation is repeated and averaged for 5 successive local maxima. The average and standard deviation of damping ratios at each structural frequency are plotted in Figure~\ref{fig:dampingratio}(a). Across all frequencies, we find that the damping ratio is over an order of magnitude larger than the structural damping ratio of $\zeta_s=0.003$ set by the parameters of the CPS. For both test objects, there is a high standard deviation in damping ratios due to damping ratio changing throughout the ringdown. Thus, the system may be experiencing nonlinear damping due to nonlinear fluid drag, wake dynamics, viscous losses and other flow characteristics.

To model and understand the nonlinear damping experienced by the whisker-inspired test objects, we fit two nonlinear damping models to our experimental data: a quadratic drag model and a Van der Pol damping model. Time series showing the predicted position and force for both damping models compared to experimental data are plotted in Figure~\ref{fig:dampingratio} (e) and (f), respectively, for the $f_s=1.1$ Hz. An estimate of the quadratic damping was made by fitting the damped amplitude response, $y(t)$ to equation:
\begin{equation}
    m \ddot{y}+b\dot{y}+ky+b_f\dot{y}|\dot{y}|=0
\label{eq:quad}
\end{equation} 
where $m, k$, and $b$ are the mass, stiffness, and structural damping enforced by the CPS (Figure~\ref{fig:set up}) while $b_f$ is the (unknown) quadratic fluid damping coefficient. The quadratic drag model is applied to account for the effects of the fluid slowing the object's motion, assuming the fluid drag is dependent on the velocity \citep{morison1950force}. Higher velocity creates a larger force opposite to the direction of motion. 
A comparison of the plots in Figure~\ref{fig:dampingratio}(a) and (c) reveals that the damping ratio and quadratic fluid damping coefficient $b_f$ follow the same trend across frequencies: the damping decreases with increasing frequency. For all tested wake frequencies, the damping experienced by the elliptical cylinder is higher than that experienced by the vibrissa with decreasing difference in value as structural frequency is increased. The vibrissa vibration is significantly less damped at lower frequencies, allowing the vibrissa to exhibit higher excitation amplitudes than the elliptical cylinder. The quadratic drag model's fit to the positional measurements does not accurately describe the physical system as it decays. The measured amplitude damps out much quicker than predicted by the quadratic drag model, and the measured frequency shifts over time. 

The Van der Pol damping model contains an amplitude dependent damping \citep{hardika2024improved}. When the amplitude is low, the fluid damping can be described by $\mu \dot{y}$, while at high amplitudes it is described by $-\mu y^2\dot{y}$. The Van der Pol damping was estimated by fitting to the equation:
\begin{equation}
    m \ddot{y}+b\dot{y}+ky+\mu(1-y^2)\dot{y}=0
\label{eq:vdp}
\end{equation} 
where $\mu$ is the (unknown) Van der Pol damping coefficient. For each ringdown experiment, we find the optimal value of $\mu$ that best fits an analytical prediction of the model to the experimental measurement. Mean and standard deviation values of $\mu$ for the repeated structural frequency tests are plotted in Figure~\ref{fig:dampingratio} (d). Note that for our system, the $\mu$ values that best fit our experimental results are all negative, meaning that, as expected, the system experiences positive damping at larger amplitudes, but negative damping at low amplitudes. This successfully captures the change in the character of the fluid damping - that it suppresses the large amplitude wake-induced-vibrations (WIV), but recovers the natural VIV instability  at low amplitues (which is balanced by the structural damping). The amplitude of $\mu$ is lower for the  vibrissa test object, in agreement with the observation that the vibrissa maintains higher amplitudes of oscillation during WIV as well as in the ringdown experiments. 
   
\begin{figure}[H]
    \includegraphics[width=\textwidth]{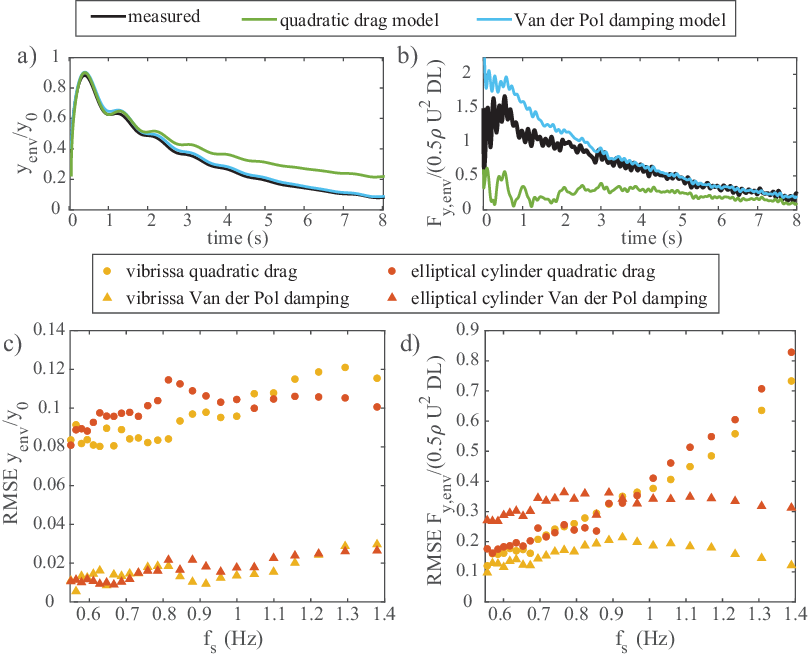}
    \centering
    \caption{\textbf{Quality of fit for damping models.} Hilbert envelopes for predicted and measured (a) position and (b) force for a vibrissa test case at structural frequency $f_s=1.1$ Hz. Root-mean-square error in each damping models' Hilbert envelopes for (c) position and (d) plotted for each tested structural frequency. RMSE of the quadratic drag model is denoted by circular markers and RMSE of the Van der Pol model is denoted by triangle marker. Colors orange and yellow represent results for the elliptical cylinder and vibrissa test objects, respectively.}
    \label{fig:envelopes}
\end{figure}

To quantitatively compare the accuracy of the tested damping models, we apply Hilbert envelops,  obtained by taking the magnitude of the time series signal and its Hilbert transform. This provides a smooth estimate of the signal’s instantaneous amplitude \citep{Thrane1985Hilbert}. Hilbert envelopes were calculated and plotted in Figure~\ref{fig:envelopes} for $f_s=1.01$ Hz for both models' predicted (a) position and (b) force. Hilbert envelopes were found for all trials and compared to the measured position and force by calculating root-mean-square error (RMSE) for both models at all tested frequencies. This is shown in Figure~\ref{fig:envelopes} (c) and (d) for position and force RMSE, respectively. Triangle markers indicate the Van der Pol damping model and circle markers indicate the quadratic drag model. Comparing the results for position first, in Figure~\ref{fig:envelopes}(c), we see across all structural frequencies, the Van der Pol damping model has significantly less error than the quadratic drag model for both the elliptical cylinder and vibrissa. This result is consistent with the Van der Pol modeling analysis done by \citet{facchinetti2004coupling} and \citet{hardika2024improved} for circular cylinder vibration. The fact that the fluid damping of both the elliptical cross-section and circular cross-section test objects can be accurately described by a Van der Pol damping term suggests that the oscillatory motion through the fluid imparts position dependent damping regardless of geometry and it is simply strength in this damping contribution that changes for different model shapes. Subsequently, there is no significant difference in model accuracy between the two reduced aspect ratio test objects when it comes to RMSE in $y$-position, nor does the error appear to hold any frequency dependence. This does change however when we evaluate the force data. As structural frequency increases, the error in the quadratic drag model for both test object increases as well. The direct positive correlation indicates that the accuracy of the quadratic drag model is heavily dependent on frequency of the system. The Van der Pol damping model has similar error across all tested frequencies, just as with the position results, demonstrating model accuracy is not frequency dependent for position or force. The RMSE in the Van der Pol damping model predicted force on the vibrissa is roughly half that predicted for the elliptical cylinder model. Thus, while the Van der Pol damping model is significantly more accurate than the quadratic damping model in modeling the fluid damping for both objects, the Van der Pol damping model describes the physical response of the  vibrissa with higher accuracy than that of the elliptical cylinder. This is suggests differences in the vortical structures shed from the differing body shapes, but this needs to be confirmed with velocity measurements. 

\begin{figure}[H]
    \includegraphics{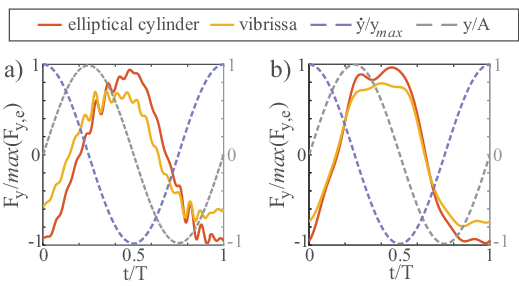}
    \centering
    \caption{\textbf{Transverse force experienced during prescribed sinusoidal heaving.} phase-averaged force experienced by the elliptical cylinder and vibrissa models during prescribed sinusoidal heaving with frequency of (a) 0.5 Hz and (b) 1.5 Hz. The dashed lines indicate heave position (gray) and velocity (purple) over the cycle.}
    \label{fig:fy_cycle}
\end{figure}

To further characterize the difference in the fluid forces experienced, we recorded the forces experienced by each test geometry subjected to \emph{prescribed} sinusoidal heave motion ($A= 3$ cm, $f=$ 0.5 - 1.5 Hz) with a mean flow ($U = $0.3 m/s). Note that that observed free motion is very close to sinusoidal (Fig~\ref{fig:free flow}) and so this mimics the natural VIV very closely. 
Figure~\ref{fig:fy_cycle} shows phase-averaged $y$-force (force in the transverse direction to the flow) of both models at (a) 0.5 Hz and (b) 1.5 Hz frequencies, with the prescribed heave displacement and velocity over a cycle plotted in gray and purple, respectively. The forces are shown scaled by the maximum $y$-force of the elliptical cylinder model, the displacement $y$ is nondimensionalized by the amplitude of the sinusoidal motion, $A$, and the velocity, $\dot y$, is scaled by the maximum velocity in the cycle. 

The force experienced by the test objects during prescribed motion corroborate the results of the ringdown experiments: The elliptical cylinder experiences higher force than the vibrissa model, and the difference in $y$-force experienced by the vibrissa compared to the elliptical cylinder is significantly larger at lower frequency. At $f = 0.5$ Hz, the root-mean-square (RMS) $y$-force experienced by the vibrissa is 35\% lower than that experienced by the elliptical cylinder, while at $f = 1.5$ Hz, the reduction is only 17\%. Other studies have found that when held stationary in a flow, the force experienced by undulated vibrissa models has an even greater reduction in lift force than then elliptical cylinder. Simulations by \citet{liu2019phase} of flow past stationary undulated and uniform span elliptical cylinders found the RMS of the lift coefficient for the vibrissa geometry was more than one order of magnitude lower than that of the elliptical cylinder and simulations by \citet{yoon2020effect} found that major-axis undulation lead to more than 89\% reduction in RMS lift coefficient of the elliptical cylinder. Even in live seals, \citet{hanke2010harbor} recorded the dynamic forces on harbor seal whiskers to be on average 6.2 times lower than the forces acting on the non-undulated sea lion whiskers (although it must be noted that the aspect ratio differed between these whiskers).

For spring-mass-damper systems that are free to move, the dominating force component is informed by the phase of the force vs. the position: A phase shift of zero degrees indicates a purely stiffness controlled system, while a ninety degree phase indicates pure damping. We calculated the phase shift between the prescribed sinusoidal heave position and the first harmonic of our force response, found by fitting the phase-averaged $y$-force to a Fourier series. The phase shift for the elliptical cylinder heaving at 1.5 Hz, and vibrissa model heaving for both 0.5 Hz and 1.5 Hz cases are all very similar (54-degrees, 51-degrees, and 52-degrees, respectively). However, the force experienced by the elliptical cylinder during the prescribed 0.5 Hz sinusoidal motion is 72-degrees out of phase with position. Thus, the effects of damping provide greater contribution to the force experienced by the elliptical cylinder at low frequencies. The significant change in phase between the position and $y$-force experienced by the elliptical cylinder at low frequency may be a product of vortical structures shed from this object that differ from those shed by the undulated vibrissa geometry. As demonstrated by the many works summarized in \citet{lekkala2022recent}, at higher effective flow velocities, where separation and boundary layer behavior is more dominated by inertia, wakes tends to appear more similar to those of classic bluff body geometries. Thus, the modifications to vortex-shedding provided by the undulated geometry of the seal vibrissa, bear more impact in reducing fluid damping during lower frequency motion. Another notable observation from Figure~\ref{fig:fy_cycle} is that the 0.5 Hz phase-averaged force curves are more sinusoidal than those in the higher frequency plot. In a perfect linear oscillator, the force response of the system can be well-defined by simple sine and cosine terms, the deviation from this indicates that the higher frequency case has a higher degree of nonlinearity. While the scope of our investigation only assessed motion with frequencies up to 1.5 Hz, where -although more similar- the vibrissa still experienced lower damping and $y$-force than the elliptical cylinder, avenues of exploration in future work may include determining at what frequency motion the responses of the smooth and undulated models converge. And further, evaluating how the wake structures shed during motion differ between frequencies and test models.  This is ongoing work in our laboratory.

\section{Conclusions}
Through characterizing the forces and displacements of the free vibration of circular cylinder, elliptical cylinder, and vibrissa models, we identified potential benefits the undulated geometry of a seal whisker presents over classical cylinder geometries. In a free flow, the reduced aspect ratio objects exhibit minimal VIV across all structural frequencies, reducing noise in sensing due to self-excitation. In the presence of a vortical wake, the reduced aspect ratio objects are excited by WIV at a frequency matching that of the upstream wake with highest amplitude oscillations occurring when structural frequency of the test object is equal to the wake frequency of the upstream foil. By evaluating two damping models, we found the vibrissa experiences lower damping than the elliptical cylinder in all tested cases, increasing its signal-to-noise ratio. The damping experienced by the objects is best described by an amplitude-dependent Van der Pol model, injecting energy into the system at low amplitudes and extracting it at high amplitudes of oscillation. Overall, we conclude that the elliptical cross-section span-wise undulated vibrissa geometry is more effective for flow sensing than smooth geometries. The lack of vibrissa excitation in an undisturbed flow makes it superior to the highly self-excited circular cylinder, while its reduced fluid damping gives it higher sensitivity to a flow disturbance than the elliptical cylinder. Our experiments found that the vibrissa experiences lower force in motion, plus higher excitation when freely vibrating than the elliptical cylinder. 
This property could be a key benefit to seals using vibrissa for predator or prey detection.


The present results provide insight into how a single vibrissa responds to a wake disturbance at a fixed angle of attack. The response to more complex wake geometries, measurements and simulations of the wake behind the vibrating structure, and the response of the body at different angles of incidence and sweep will deepen our understanding of the effects of undulation on these fluid-structure interactions, and results in these areas of study are starting to be reported \cite[e.g.][]{dunt2025:SweepAngleEffects}. Lastly, the true sensory systems of seals are made up of many vibrissae positioned throughout the face. How the vibrational information gathered from each individual sensor comes together to form understanding of a complex underwater environment is of great interest and provides direction for future studies. Experiments,  simulations and the use of deep-learning models to further investigate the interplay of multiple whisker vibrations and the ability of these networks to provide spatial maps of the surrounding flow will lead to advancements in this arena and are already being explored \cite{bodaghi2023deciphering, geng2025biomimetic}. 





\section*{Acknowledgments}
This research was supported by ONR N00014-21-1-2816. EE is supported by the NSF Graduate Research Fellowship Program.

\raggedright
\bibliographystyle{elsarticle-harv}
\bibliography{refs.bib}
\end{document}